\documentclass[11pt,preprint]{aastex}

\newcommand\rg{r_{\mathrm{g}}}
\newcommand\rgmax{r_{\mathrm{g,max}}}

\newcount\listnorom
\newcommand\listromanDE{\global\advance \listnorom by 1 
{\lowercase\expandafter{\ (\romannumeral\listnorom)}\ }}
\newcommand\newlistroman{\listnorom=0}
\newcommand\GGG{G347.3$-$0.5}
\newcommand\TBn{\Theta_{\mathrm{Bn}}}
\newcommand\dMdt{dM_{\mathrm{loss}}/dt}
\newcommand\vwind{v_w}
\newcommand\fvol{f_{\mathrm{vol}}}
\newcommand\dSNR{D_{\mathrm{snr}}}
\newcommand\EmisVol{V_{\mathrm{emis}}}
\newcommand\tSNR{t_{\mathrm{snr}}}
\newcommand\DelTime{\Delta t_{\mathrm{sh}}}
\newcommand\ProDenUpS{n_{p0}}
\newcommand\ShellDen{n_{\mathrm{shell}}}
\newcommand\ShellRho{\rho_{\mathrm{shell}}}
\newcommand\EnSN{E_{\mathrm{sn}}}
\newcommand\fHe{f_{\alpha}}
\newcommand\Mej{M_{\mathrm{ej}}}
\newcommand\Msun{M_{\odot}}

\newcommand\epRatio{(e/p)_{\mathrm{rel}}}
\newcommand\TempPro{T_{p0}}
\newcommand\TempElec{T_{e0}}
\newcommand\TempHe{T_{\alpha0}}
\newcommand\TempRatio{T_{\mathrm{e2}}/T_{\mathrm{p2}}}

\newcommand\Rsk{R_{\mathrm{sk}}}
\newcommand\Rshell{R_{\mathrm{in}}}
\newcommand\EmaxPro{E_{\mathrm{max,p}}}
\newcommand\EmaxElec{E_{\mathrm{max,e}}}
\newcommand\MAZ{M_\mathrm{A0}}
\newcommand\MSZ{M_\mathrm{S0}}
\newcommand\Rsub{r_\mathrm{sub}}
\newcommand\DStemp{T_{\mathrm{p2}}}
\newcommand\DStp{T_{\mathrm{tp}}}
\newcommand\EffRel{\epsilon_{\mathrm{rel}}}
\newcommand\Emax{E_{\mathrm{max}}}

\newcommand{\xx}[1]{\!\times\!10^{#1}}
\newcommand\gameff{\gamma_\mathrm{eff}}
\newcommand\Frel{F_\mathrm{rel}}
\newcommand\Fesc{F_\mathrm{esc}}
\newcommand\itt{ }
\newcommand\bff{ }

\newcommand\RH{Rankine-Hugoniot}

\newcommand\kmps{km s$^{-1}$}

\newcommand\Rtot{r_\mathrm{tot}}
\newcommand\Vsk{V_\mathrm{sk}}
\newcommand\etainjP{\eta_\mathrm{inj,p}}
\newcommand\etamfp{\eta_\mathrm{mfp}}
\newcommand\pmax{p_\mathrm{max}}
\newcommand\pcc{cm$^{-3}$}
\newcommand\muG{$\mu$G}
\newcommand\iec{i.e.,}
\newcommand\egc{e.g.,}
\newcommand\etal{et al.}
\newcommand\syn{synchrotron}

\newcommand\brem{bremsstrahlung}

\newcommand\IC{inverse-Compton}
\newcommand\pion{pion-decay}

\newcommand\alf{Alfv\'en}

\shorttitle{Observations and Modeling of SNR G347.3-0.5}
\shortauthors{Ellison et al.}

\begin{document}     
   
\title{Broad-band Observations and Modeling of the
Shell-Type Supernova Remnant G347.3-0.5}

\author{
Donald C.~Ellison\altaffilmark{1},
Patrick Slane\altaffilmark{2}, and
Bryan M.~Gaensler\altaffilmark{3,4}
}

\altaffiltext{1}{Department of Physics, North Carolina State
 University, Box 8202, Raleigh NC 27695, U.S.A.; E-mail:
 don\_ellison@ncsu.edu}
\altaffiltext{2}{Harvard Smithsonian Center for Astrophysics, 60
 Garden Street, Cambridge, MA 02138, U.S.A.; E-mail:
slane@cfa.harvard.edu}
\altaffiltext{3}{Hubble Fellow}
\altaffiltext{4}{Center for Space Research, Massachusetts Institute
 for Technology, Cambridge, MA 02139, U.S.A.; E-mail:
bmg@space.mit.edu}

\slugcomment{ApJ, in press August 2001 (Submitted March 2001)}

\begin{abstract}
The supernova remnant G347.3--0.5 emits a featureless power-law in
X-rays, thought to indicate shock-acceleration of electrons to high
energies. We here produce a broad-band spectrum of 
the bright NW limb of
this source by combining radio observations from the Australia
Telescope Compact Array (ATCA), X-ray observations from the Advanced
Satellite for Cosmology and Astrophysics (ASCA), and TeV $\gamma$-ray
observations from the CANGAROO imaging \v{C}erenkov telescope.
We assume this emission is produced by an electron population
generated by diffusive shock acceleration at the remnant
forward shock.  The nonlinear aspects of the particle acceleration
force a connection between the widely different wavelength bands and
between the electrons and the unseen ions, presumably accelerated
simultaneously with the electrons.
This allows us to infer the relativistic proton spectrum and
estimate ambient parameters such as the supernova explosion energy,
magnetic field, matter density
in the emission region, 
and efficiency of the shock
acceleration process.  We find convincing evidence that the shock
acceleration is efficient, placing $>25$\% of the shock kinetic energy
flux into relativistic ions.
Despite this high efficiency, the maximum electron and proton
energies, while depending somewhat on assumptions for the compression
of the magnetic field in the shock, are well below the observed `knee'
at $\sim 10^{15}$ eV in the Galactic cosmic-ray spectrum.
\end{abstract}        


\section{INTRODUCTION}

The supernova remnant G347.3--0.5 (ROSAT catalog source name: RX
J1713.7--3946) is one of a growing number of supernova remnants (SNRs)
showing nonthermal X-ray emission attributed to shock accelerated
electrons.\footnote{As of this writing, the list of SNRs with
nonthermal X-ray emission associated with shock acceleration rather
than, or in addition to, pulsar-powered emission includes: Tycho,
Kepler, Cas A, SN1006, G266.2-1.2, G347.3-0.5, RCW 86, G156.2+5.7, and
3C397.}
In some cases, the nonthermal emission is evidenced by continuum
emission extending to $\sim$10~keV or above, usually superimposed with
weak lines, or,
as with
SN1006, G347.3-0.5, and G266.2-1.2, by a complete lack of line
emission in the observed X-ray band.\footnote{SN1006 shows line
emission from the central regions, but nearly featureless spectra from
the bright rim \citep[\egc][]{Koyama95}.}  In all cases where the
emission is not associated with a compact object, the nonthermal
X-rays have been interpreted as \syn\ emission from shock accelerated
TeV electrons \citep[\egc][]{MdeJ96,Allen97,Tanimori98,RK99}.  If this
is true, the existence of these TeV electrons has important
implications, not only for the production of cosmic rays, but also for
the thermal properties of the shock heated, X-ray emitting gas in
SNRs.

Here we compare broad-band observations of G347.3--0.5 with results
from a model 
combining hydrodynamic simulations of SNR evolution
\citep[\egc][and references therein]{BE2001} with 
% of 
nonlinear diffusive particle acceleration 
% in the remnant blast wave 
\citep[\egc][]{EBB00}. The observations combine new
radio data from the Australia Telescope Compact Array (ATCA) with
existing X-ray \citep{Slane99} and $\gamma$-ray data
\citep{Muraishi2000}.
Our broad-band comparison of observations and model fits 
%% over $\sim 18$ decades in energy 
is a powerful way to constrain model parameters,
but uncertainties from directly comparing radio, X-ray, and
$\gamma$-ray observations are unavoidable and should be kept in mind.
In an attempt to limit errors from uncertainties in emission volumes,
we only consider observations from the northwest portion of \GGG\
where the center of the TeV $\gamma$-ray response function coincides
with the brightest radio and X-ray emission. 

Previous work \citep[\egc][]{Reynolds98} indicates that the steepness
of the nonthermal X-ray \syn\ spectrum in shell-like SNRs results
because these photons are generated by electrons in the steeply
falling, high-energy tail of the distribution.  The actual maximum
electron energy, however, depends on the magnetic field strength,
$B_2$, in the emission region. If these electrons are accelerated by
the first-order Fermi shock acceleration mechanism as we assume,
it is almost certain that the same shocks accelerate ions, but no
clear evidence of superthermal ions has yet been obtained (i.e., no
\pion\ feature has been unambiguously identified at $\sim 100$ MeV in
SNRs). Until a \pion\ feature is observed, all information on the
energetic ions must be obtained through modeling. In shocks
accelerating particles efficiently,
the unseen ions dominate the shock dynamics and largely control the
electron emission features \citep[\egc][]{Drury83,JE91}.

Our main results are: 
(i) A spherically symmetric, wind-shell model 
with efficient acceleration
%homogeneous model of nonlinear shock acceleration 
can produce broad-band continuum spectra in
excellent agreement with the observations of G347.3-0.5;
(ii) Test-particle models of \GGG\ (\iec\ ones where less than a
few percent of the total energy ends up in relativistic particles) can
be excluded with good confidence;
% also fit the 
% {\it shape}
%  of the broad-band emission reasonably well, but under-produce the
% flux by a large factor. This and other considerations lead us to
% conclude that \GGG\ is currently accelerating particles efficiently;
%
(iii) The maximum electron energy in \GGG\ is well below $10^{14}$ eV.
Thus, \GGG\ joins several other SNRs now known {\it not} to produce
electrons above $10^{14}$ eV \citep[\egc][]{RK99}, well below the
so-called knee in the Galactic cosmic-ray spectrum at $\sim 10^{15}$
eV.  
The maximum proton energy depends somewhat on whether or not the
ambient magnetic field is compressed by the shock. If the field is
compressed as the density, the maximum proton energy
is approximately the same
as the electron $\Emax$.
In the other extreme, where no compression of the field
occurs, we find that G347.3-0.5 could
currently be producing
protons to $\sim 70$ TeV and Fe$^{+26}$ to $\sim 10^{15}$ eV.
Unless substantial magnetic field amplification occurs, \GGG\
will not produce higher energy protons as it ages; and
(iv) Our main conclusions concerning 
maximum particle energies and acceleration efficiency,
are relatively insensitive to uncertainties stemming from
the morphology of \GGG.
% do not depend substantially on whether or not this SNR is currently
% interacting with a molecular cloud \citep[see][]{Slane99}.

There are strong arguments for a SNR origin of cosmic rays based
mainly on ion composition \citep[\egc][]{EDM97,MDE97} and total energy
requirements 
\citep[\egc][]{Axford81,BE87,DMV89}. 
The observations of nonthermal
X-ray emission showing that TeV electrons are produced in SNRs adds
support for this argument.  However, there is still no direct evidence
that ions are generated in 
particular
SNRs in a fashion consistent with cosmic-ray observations. 
The low maximum proton energy inferred for G347.3-0.5 may present a
problem for the scenario that SNRs are the main sources of Galactic
cosmic rays below the knee and suggest that the cosmic-ray proton
component above $\sim 10^{14}$ eV might come from some subset of SNRs
substantially different from \GGG. Alternatively, young SNRs which
only produce $\sim 10^{13}$ eV particles when they are bright in
X-rays, may somehow be able to produce higher energy particles at
later times, perhaps by amplifying the magnetic field
\citep[\egc][]{Keohane98,JJ99,Lucek00}. We note that the high magnetic
fields inferred for Cas A make it likely that this particular remnant
{\it does} produce $\sim 10^{15}$ eV protons \citep[\egc][]{Allen97}.

\section{Observations}

\begin{figure}[!hbtp]              % Figure 1
\epsscale{0.6}
\plotone{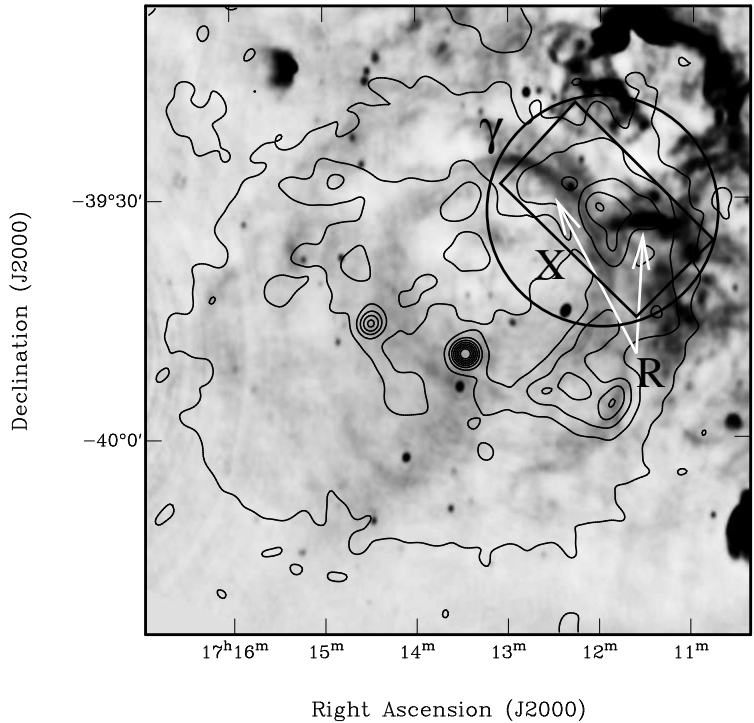}
\figcaption{
G347.3-0.5 at 1.36~GHz, from the ATCA, with X-ray contours from the ROSAT 
PSPC \citep{PA96}. 
The modeling presented here corresponds to emission from the northwest 
limb of the SNR. The radio flux used for our modeling
is from the 
two bright filaments indicated with white arrows. 
The X-ray spectrum was extracted from the region indicated 
by a box. 
The circle on the figure is placed at the centroid of the 
$\gamma$-ray emission detected by the CANGAROO telescope; the radius of the 
circle corresponds to the telescope point response function.
\label{fig:G347image}}
\end{figure}

G347.3$-$0.5 was observed with the {\it Advanced Satellite for
Cosmology and Astrophysics} (ASCA) on 1997 March 25 for a total of
50~ks using three pointings of 10~ks duration and one (in the fainter
southeast region) of 20~ks duration \citep{Slane99}.  Unlike the
line-dominated X-ray spectra typical of SNRs, the spectrum for
G347.3$-$0.5 is observed to be featureless and to extend to at least
10~keV, where the ASCA effective area becomes very small.  A power law
spectrum provides an excellent fit to the X-ray data, with some
indication that the spectral index varies between the brighter shell
regions and the more diffuse interior and eastern regions.
The brightest X-ray emission occurs along the northwest limb of the
SNR (see Figure~\ref{fig:G347image}), where there is some indication
that the remnant may be interacting with a molecular cloud or a
stellar wind bubble shell, although the low inferred density for this
region more likely indicates some density enhancement caused by a
perturbation of the pre-SN wind.
The unabsorbed flux density from this
region of the remnant is $F_x(0.5-10 \, \mbox{keV}) = 1.6 \times
10^{-10}\, \mbox{erg cm}^{-2} \mbox{s}^{-1}$ with a power law photon
index of $2.41^{+0.05}_{-0.04}.$

Described here are new radio observations of G347.3--0.5 made with the
Australia Telescope Compact Array (ATCA), a six-element synthesis
telescope located near Narrabri, NSW, Australia.  Observations at
1.36~GHz were carried out in January, March, and April 1998, in each
case in a different array configuration to give maximal $u-v$
coverage. The field was imaged in a 10-point mosaic, to ensure that
the entire extent of the SNR, as well as nearby bright sources, were
in the field-of-view.
The interferometric visibilities were edited to remove corrupted data
and interference. 
Calibrations were then applied to solve for the time-variable
atmospheric gains above each antenna and for the polarization
leakage in the receivers. An absolute flux density scale
was applied to the data using observations of PKS~B1934--638,
for which a flux density of 15.0~Jy was assumed at a frequency
of 1.4~GHz.
%
%Calibrations were then applied to solve for the
%absolute flux density scale, time-variable atmospheric gains above each
%antenna, and polarization leakage in the receivers. 
The visibilities
from all 10 pointings were then inverted to produce a single image,
which was deconvolved using the mosaiced maximum entropy approach
described by \citet{Sault96}. The final image
was then produced by applying a Gaussian restoring beam of dimensions
$46''\times 36''$. 
The sum of the 
flux from the two bright filaments in the northwest region of the
remnant \citep[indicated in Figure~\ref{fig:G347image} and labeled as
arc 1 and 2 in][]{Slane99}, corresponding to that from which the X-ray
flux above was measured, is $S(1.36 \, \mbox{GHz}) = 4 \pm 1 \,
\mbox{Jy}$.
As noted in \citet{Slane99}, while arc 2 (the right-most arc in
Figure~~\ref{fig:G347image}) is askew from the perceived SNR shell, it
does overlap an X-ray filament. Arcs 1 and 2 have approximately the
same radio flux.

G347.3-0.5 has also been detected at TeV energies with the CANGAROO 3.8m
atmospheric imaging \v{C}erenkov telescope \citep{Muraishi2000}.  As
with the X-ray and radio observations, the $\sim 5.6\sigma$ detection
is from the NW rim of the remnant 
(indicated with a circle in Figure~\ref{fig:G347image})
and has an estimated flux of $5.3
\pm 0.9 \mbox{[statistical]} \pm 1.6 \mbox{[systematic]} \times
10^{-12}$ photons cm$^{-2}$ s$^{-1}$ at energies $\ge 1.8 \pm 0.9$
TeV.

The fluxes from these three detections are shown in
Figures~\ref{fig:broadband} and \ref{fig:vary} where we have increased
the observational error bars on the radio and TeV points by a factor
of two (light-weight bars) to compensate for uncertainties in the
relative emission volumes.
In particular, the increased radio error bar will accommodate the
possibility that arc 2 is not associated with the SNR shell and thus
none of our results will change if this is the case. 

\section{Model}

\subsection{Nonlinear diffusive shock acceleration}

Collisionless shocks are known to accelerate particles efficiently. In
the heliosphere, where shocks are observed directly with spacecraft,
there is clear evidence that the quasi-parallel earth bow shock (with
a sonic Mach number, $\MSZ < 10$) can place 10-30\% of the solar wind
kinetic energy flux into superthermal particles
\citep[\egc][]{EMP90}. Interplanetary shocks (IPSs) also accelerate
ambient particles but generally with lower efficiencies due to their
lower Mach numbers (generally $\MSZ \lesssim 3$ for IPSs)
\citep[\egc][]{Baring97}.
However, on at least two occasions, exceptionally strong IPSs have been
observed to accelerate particles with high efficiency
\citep{Eichler81,Terasawa99}.
Hybrid plasma simulations showing direct
injection and acceleration of ambient particles are consistent with
these observations \citep[\egc][]{STK92}, as are convection-diffusion
models \citep[\egc][]{KJ97}.

Compared to shocks in the heliosphere, it might be presumed that the
strong forward and reverse shocks in young SNRs are much more
efficient.  As far as we can tell, there are no important differences
in typical plasma parameters (\iec\ density, magnetic field,
temperature, composition) between the solar wind and the ambient
ISM. The only real differences come from the much larger shock spatial
and time scales and much higher flow speeds and Mach numbers in SNRs.
In general, these factors should increase the acceleration efficiency
substantially over heliospheric shocks, but some qualifications must
be made.

At quasi-parallel shocks (those where the upstream magnetic field
makes an angle, $\TBn$, of less than about $45^\circ$ with the shock
normal), injection from the shocked heated plasma should occur easily
and the fraction of incoming ram kinetic energy flux going into
superthermal particles increases with Mach number and shock size
and/or age (as measured in particle gyroradii or gyroperiods).  If
however, the fraction of shocked heated particles injected is small
enough, high Mach number, test-particle solutions can result
\citep[see][]{BE99}. In quasi-perpendicular
shocks (ones where $45 < \TBn \lesssim 90^\circ$), injection and
acceleration of shock heated thermal particles 
is suppressed 
above some Mach number unless strong turbulence is present
\citep[see][for example]{EBJ96}. Highly oblique interplanetary shocks
are observed to accelerate the ambient solar wind \citep[even without
pickup ions,][]{EJB99}, but these are low Mach number shocks. As the
Mach number increases, injection
becomes more likely
if the turbulence is strong enough for cross-field diffusion in the
downstream plasma to be important.

While the shock obliquity will surely vary around a SNR and some
portions of the shocks may be quasi-perpendicular with low injection
rates, other regions should be parallel enough for efficient injection
and acceleration to occur. Radio emission is a sure sign that
relativistic electrons are present and flux levels, sharp radio
edges, etc., show
that these particles are locally accelerated  in young SNRs
rather than just
compressed Galactic cosmic-ray electrons \citep[\egc][]{ABR94}. 
The shock acceleration and evolutionary models we use
here ignore effects from oblique magnetic fields, other than those
that are mimicked by a low injection efficiency. We do assume,
however, that the magnetic field responsible for \syn\ radiation can
be compressed by the shock.
For a recent discussion of how the injection efficiency may vary with
$\TBn$ in SNRs, see \citet{Volk2001}.

The nonlinear effects in efficient
shock acceleration are of three basic kinds:
(i) the self-generation of magnetic turbulence by accelerated
particles, (ii) the modification (\iec\ smoothing) of the shock
structure by the backpressure of accelerated particles, and (iii) the
increase in shock compression ratio as relativistic particles are
produced and some high-energy particles escape from the system.
Briefly, (i) occurs when counter-streaming accelerated particles
produce turbulence in the upstream magnetic field which amplifies as
it is convected through the shock. This amplified turbulence results
in stronger scattering of the particles, and hence to more
acceleration, quickly leading to saturated magnetic turbulence levels
near $\delta B/B \sim 1$ in strong shocks.
This is the so-called Bohm limit where the mean free path is of the
order of the gyroradius, and in this limit, the distinction between
parallel and oblique shocks blurs.  The wave-particle interactions
produce heating in the shock precursor which may be observable and
which lowers the overall acceleration efficiency.
Effect (ii) occurs because the accelerated population presses on the
upstream plasma and slows it.  An upstream precursor forms, in which
the flow speed (in the shock frame and in the absence of
instabilities) is monotonically decreasing.  Since particle diffusion
lengths are increasing functions of momentum, high momentum particles
sample a broader portion of the flow velocity profile, and hence
experience larger effective compression ratios than low momentum
particles.  Consequently, higher momentum particles have a flatter
power-law index than those at lower momenta (producing a concave
upward spectral curvature) and can dominate the pressure in a
non-linear fashion.
We call the ratio of densities spanning the entire upstream precursor
the overall compression ratio, $\Rtot$, to distinguish it from the
small scale subshock compression ratio, $\Rsub$, where most of the
shock heating occurs.  For a more complete discussion of the nonlinear
effects of diffusive shock acceleration, see \citet{BE99}.

The conservation of mass, momentum, and energy fluxes, i.e., the \RH\
relations, yield the standard expression for the compression ratio
in shocks where the magnetic field is not dynamically important
and an insignificant fraction of energy is placed in superthermal
particles (i.e., test-particle shocks):
\begin{equation}
\Rtot =
\frac{(\gamma + 1) \MSZ^2}{(\gamma - 1) \MSZ^2 + 2}
\ ,
\end{equation}
where $\MSZ = \sqrt{\rho_0 \, \Vsk^2 / (\gamma \, P_0)}$ 
is the sonic Mach number, $\gamma$ is the ratio of
specific heats, $\rho_0$ is the unshocked density, 
$P_0$ the unshocked pressure,
and $\Vsk$ is the
shock speed.\footnote{Here and
elsewhere, the subscript `0' indicates unshocked values and the
subscript `2' indicates shocked values.}
In test-particle shocks, $\Rtot = \Rsub$.
The ratio of shocked to unshocked pressure follows:
\begin{equation}
\frac{P_2}{P_0} = \frac{2 \gamma \MSZ^2 - (\gamma - 1)}{\gamma + 1}
\ .
\end{equation}
If $\MSZ \gg 1$,
\begin{equation}
\Rtot \simeq \frac{\gamma +1}{\gamma - 1}
\ ,
\end{equation}
and
\begin{equation}
P_2 \simeq P_0 \, \frac{2 \gamma \MSZ^2}{\gamma + 1} =
\frac{2 \rho_0 \Vsk^2}{\gamma + 1}
\ .
\end{equation}
For a electron-proton plasma where the downstream temperatures are
equal, the post-shock proton temperature in this test-particle limit,
$\DStp$, is:
\begin{equation}
\label{eq:TpTemp}
\DStp \simeq
\frac{1}{\Rtot} 
\frac{m_p \Vsk^2}{k (\gamma + 1)}
\ ,
\end{equation}
where $k$ is Boltzmann's constant and $m_p$ is the proton mass.

Young SNR shocks are expected to have high Mach numbers and if the
test-particle situation is assumed with $\gamma \simeq 5/3$, we expect
$\Rtot \simeq 4$. However, when shocks accelerate particles the
overall compression ratio increases above test-particle values (even
in non-radiative shocks) for two reasons (effect iii above).
First, as relativistic particles are produced and contribute
significantly to the total pressure, their softer equation of state
makes the shocked plasma more compressible
($\gamma  \rightarrow 4/3$).  
Second, and most important, is that particles escape from strong
shocks draining energy
flux which must be compensated for
by ramping up the overall compression ratio to conserve the
fluxes.\footnote{As long as the speeds of the escaping particles are
much greater than the shock speed, escaping momentum and mass fluxes
can be neglected \citep{Ellison85}.
}
 Just as in radiative shocks, this is equivalent to $\gamma
\rightarrow 1$.
For strong injection, it
can be shown 
\citep[][]{BE99} that $\Rtot$ is an ever increasing function of Mach
number, \iec\
\begin{equation}
\label{eq:machsonic}
\Rtot \simeq \cases{
 1.3 \, \MSZ^{3/4}
 &if \ \  $1 \ll \MSZ^2 \ll \MAZ$
\ ;
\cr
\noalign{\kern5pt}
 1.5\, M_{A0}^{3/8}
&if \ \ $1 \ll \MAZ < \MSZ^2$
\ ,
\cr
}
\end{equation}
where $\MAZ$ is the \alf\ Mach number \citep[see
also,][]{KE86,Malkov97}.  Simultaneously with $\Rtot$ increasing, the
subshock compression ratio, $\Rsub$, which is mainly responsible for
heating the gas, must drop below the test-particle value. Thus,
high compression ratios are accompanied by low post-shock
temperatures (in efficient shocks, the post-shock proton temperature
can easily be $1/10 \, \DStp$).\footnote{ \citet{Hughes00b} have used
this effect to reconcile the high values of the shock speed determined
from expansion measurements and the low post-shock electron
temperature determined from X-ray line models in SNR E0102.2-7219.}
If injection is weak enough, however, equation~(\ref{eq:machsonic})
does not apply and high Mach number shocks do not
accelerate enough particles to cause the total compression ratio to
increase above $\sim 4$ \citep[see][for the parameters describing this
condition]{BE99}. 

The most important parameters associated with nonlinear shock
acceleration are the Mach numbers (\iec\ the shock speed, $\Vsk$,
pre-shock hydrogen number density, and preshock magnetic field,
$B_0$),
the injection efficiency, $\etainjP$ (\iec\ the fraction of total
protons which end up with superthermal energies), and the maximum
proton energy produced, $\EmaxPro$.  
As described in \citet{BE99}, our model includes \alf\ heating in the
precursor which reduces the efficiency compared to adiabatic heating.

We refer the reader to \citet{EBB00} for a discussion of our treatment
of electron injection and acceleration.

\begin{figure}[!hbtp]              % Figure 2
\epsscale{0.6}
\plotone{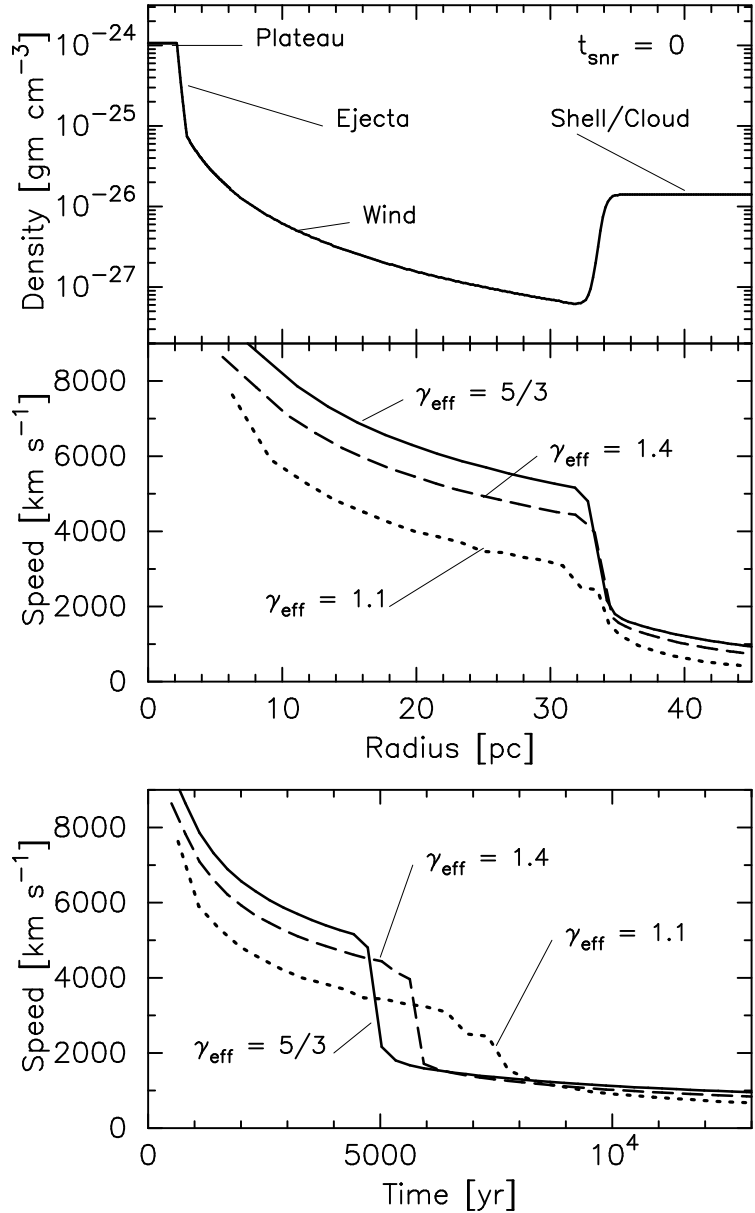}
\figcaption{
The top panel shows a snapshot of a typical density profile at the
start of the hydrodynamic simulation (i.e., at $\tSNR = 0$). The
spatial scale assumes that \GGG\ has a diameter of $\sim 45'$ and is 6
kpc away. 
The inner edge of the shell is at $\sim
34$ pc and the current radius of the forward shock is at $\sim 40$ pc.
%
%\DCE\DCE
The middle panel shows the evolution of the forward shock speed as a
function of radius and the bottom panel shows the shock speed as a
function of remnant age. The different curves in the lower two panels
were obtained by running the hydro code with the indicated effective
ratios of specific heats.  In all cases, the speed drops sharply as
the forward shock enters the shell.
\label{fig:preSN}}
\end{figure}

\subsection{SNR Evolution and Model Parameters}

\citet{Slane99} suggest that the forward shock of \GGG\
is still within the pre-SN stellar wind bubble with part of it
interacting with denser material.
We describe this evolution with a spherically symmetric, wind-shell
model with a pre-SN density distribution as shown in the top panel of
Figure~\ref{fig:preSN}.
The initial density profile in the ejecta has a power law
distribution, $\rho \propto r^{-n}$, with a constant density plateau
at small radii, as described in \citet{Chev82a}.  Since we only
consider the forward shock at fairly late stages in the SNR evolution,
our results are insensitive to $n$ and the position of the plateau.
They are also insensitive to whether a power-law or exponential ejecta
mass density distribution \citep[\egc][]{dwa98} is assumed. For
concreteness, we use $n=9$ in all of our models.
%, and
%the mass of the ejecta.
%
As in \citet{Chev82b}, the density of the pre-SN wind is given by 
\begin{equation}
\label{eq:wind}
\rho_w = A_0 \, r^{-2}
\ ,
\end{equation}
where $A_0 = \dMdt / (4 \pi \, \vwind)$, 
$\dMdt$ is the pre-SN mass loss rate, and $\vwind$ is the constant
wind speed.
%
%$s=0$ models a uniform ambient medium, while $s=2$ models a
%steady pre-supernova wind.  If $s=0$, $A_0$ in equation~(\ref{eq:wind})
%is the ambient density, \iec\
%
%$A_0 = \rho_0 = m_p \, \ProDenUpS \, (1 + 4\, \fHe)$, where $\fHe=0.1$
%is the number fraction of helium in the ambient gas.
%
%If $s=2$,
%%
%\begin{equation}
%A_0 = \frac{\dMdt}{4 \pi \, \vwind}
%\ ,
%\end{equation}
%
%where $\dMdt$ is the presupernova mass loss rate and $\vwind$ is the
%wind speed.
%
At some distance from the explosion, we place a shell of uniform
density, $\ShellRho = m_p \, \ShellDen \, (1 + 4\, \fHe)$ --- our
spherically symmetric representation of the density enhancement
suggested by \citet{Slane99}. Here,
$\fHe=0.1$ is the number fraction of helium and $\ShellDen$ is the
proton number density in the shell.
While the radio image of \GGG\ suggests a shell with the strongest
emission coming from the NW rim, the complex morphology makes an exact
determination of the diameter somewhat ambiguous. Some radio and X-ray
emission extends across $\sim 60'$, but a smaller
$\sim 30'$ diameter ring-like structure is also present.  As a
compromise, we adopt an angular diameter of $45'$, and to further limit
parameters, we perform all of our fits assuming a
distance of $\dSNR = 6$ kpc \citep{Slane99}.  
With these assumptions, the outer shock is at a radius $\Rsk \sim
40$ pc and we take (somewhat arbitrarily) the inner radius of the
shell to be $\Rshell \sim 34$ pc, giving a shell thickness
approximately equal to the thickness of radio arc 1, \iec\ $(\Rsk -
\Rshell)/\Rsk \sim 3'/22.5'$.  Fortunately, none of our conclusions
depend critically on these particular values.  Furthermore, as long as
$\rho_w(\Rshell) \ll \ShellRho$, our results are insensitive to $A_0$
and we use
% $\dMdt = 2\xx{-5} \Msun$ yr$^{-1}$ and 
$A_0 = 6\xx{12}$ gm cm$^{-1}$ which is consistent with a wind
extending to 34 pc.\footnote{For $\dMdt=7\xx{-6}$ $\Msun$ yr$^{-1}$
and $v_w = 60$ \kmps\ (giving $A_0 = 6\xx{12}$ gm cm$^{-1}$), $\sim 4
\Msun$ would exist in a spherical wind extending to 34 pc.  With this
$v_w$, the wind bubble would have taken $\sim 6\xx{5}$ yr to form and
could still not have stagnated in a low density interstellar medium.
}

The evolution is calculated with the Virginia Hydrodynamics (VH-1)
time-dependent hydrodynamic numerical code in one dimension where it
is implicitly assumed that dynamic effects from magnetic fields are
negligible. For a description of this code, see \citet{BE2001} and
references therein.  
With the shock speed and other parameters obtained with the hydro
code, we calculate the electron, proton, and helium spectra and with
them the \syn, \brem, \IC, and \pion\ photon continuum spectra. We
then compare the photon emission to the observations and vary
parameters until a satisfactory fit is obtained.

\subsubsection{Approximations of the wind-shell model}

Of necessity, our model is approximate in some important ways. First,
we assume the SNR is spherically symmetric when, in fact, the emission
is strongest from one section of the rim where the shock is likely
interacting with denser material.  We compensate for this at the most
basic level by defining the emission volume, $\EmisVol$, as a
fraction, $\fvol$, of the shell volume encompassed by the forward
shock, \iec\
\begin{equation}
\label{eq:Emission}
\EmisVol = 
\fvol \,
\left ( \frac{4 \pi}{3} \right )
\left ( \, \Rsk^3 - \Rshell^3 \right )
\ .
\end{equation}
From Figure~\ref{fig:G347image} we estimate that
$0.05 < \fvol < 0.25$.

Second, we use a one-zone model where the particle spectra and photon
emission are assumed to be generated by a shock of constant speed
during the time the forward shock is in the shell. 
We ignore any evolution of the spectra or the shock during the time
the shock is in the shell and the shock speed used for acceleration is
the average speed in the shell as determined by the hydro code.
Furthermore, we neglect any contributions from the reverse shock or
reflected shocks which occur when the forward shock enters the shell.
These should be excellent approximations since the forward shock speed
doesn't vary greatly while in the shell (middle panel
Figure~\ref{fig:preSN}), particles produced earlier in the low density
wind or ejecta have been reduced by adiabatic losses, and reflected
shocks in the shell are weak compared to the forward shock.

The evolution of the SNR will be influenced by the acceleration
process through the increased compression ratios of the forward and
reverse shocks and by the escape of high energy cosmic rays.  
We have
approximated this effect by performing our hydrodynamic calculations
with 
%
%\DCE\DCE
values of the effective ratio of specific heats, $\gameff \le 5/3$, 
%
%reduced values of the ratio of specific heats, $\gamma$, 
as
described in detail in \citet{BE2001}.  The effect on the forward shock
speed is shown for typical parameters in the bottom panel of
Figure~\ref{fig:preSN}. The solid curve is for $\gameff=5/3$ ($\Rtot
\simeq4$), the dashed curve is for $\gameff= 1.4$ ($\Rtot \simeq6$), and
the dotted curve is for $\gameff=1.1$ ($\Rtot \simeq 20$). 
When 
$\gameff=5/3$, the shock reaches 40 pc after $\tSNR \sim 1.3\xx{4}$ yr,
for 
$\gameff=1.4$, it reaches 40 pc after $\sim 1.45\xx{4}$ yr,
and for $\gameff=1.1$, 40 pc is reached after $\sim 1.8\xx{4}$ yr.
While globally changing $\gameff$ only approximates the effects of
particle acceleration on the evolution (particle escape is not
explicitly included and the evolution of $\gameff$ is not modeled, for
example), we do not believe errors from this approximation are
significant since all of the models we discuss below have compression
ratios less than $\sim 6$ when the forward shock is in the shell.  
As indicated in Figure~\ref{fig:preSN}, at $\Rsk = 40$ pc $\Vsk$
varies by less than 25\% for $\Rtot \le 6$.  Despite this small change
in $\Vsk$, the particle spectra produced by a shock with $\Rtot=6$ are
still strongly nonlinear and differs importantly from test-particle
spectra as shown in the fits below.

Finally, our model ignores X-ray {\it line} emission which, for a
thermal plasma, can be much more intense than the pure bremsstrahlung
component. For example, the 0.2--10 keV flux for a $k T = 0.2$ keV
Raymond-Smith model \citep{Raymond77} is $\sim 20$ times higher than
for a thermal bremsstrahlung model with the same temperature and
emission measure. This discrepancy decreases with temperature,
dropping to $\sim 4$ at $k T = 1$ keV and $\sim 1.4$ at $k T = 2.5$ keV. 
% but is still a factor of about  
If the featureless X-ray spectrum
in \GGG\ is \syn\ emission, the underlying \brem\
continuum which we calculate must be low enough so lines, had they
been calculated, would not appear.

\subsubsection{Model parameters}

\newlistroman

Most of the parameters used in our models are 
% explained in detail 
described
in \citet{EBB00} and/or shown here in Table~\ref{table-1}. Briefly:
\listromanDE
The explosion energy, $\EnSN$, largely determines the overall dynamics
and is typically assumed to be near $10^{51}$ erg.
The ejecta mass, $\Mej$, is unimportant for the ages and
parameters we consider as long as it is less than about 3 $\Msun$. If
$\Mej \gtrsim 3 \Msun$, the reflected shock produced when the forward shock
enters the shell can interact with the density peak behind the reverse
shock producing complicated flow structures. To avoid these
complications, we limit consideration to $\Mej < 3 \Msun$ and do not
list this parameter in the Table.
\listromanDE
The magnetic field affects the
particle acceleration and largely determines the shape of the
broad-band photon emission through its influence on \syn\ emission.
The field morphology is certain to be complex in any real SNR
and will vary with location \citep[see][for an excellent 2-D MHD study
of shock-cloud interactions]{JJ99}. To consider a range of
possibilities in our idealized, spherically symmetric model, we assume
the downstream field, $B_2$, is either equal to the upstream
(unshocked) field, i.e., $B_2 = B_0$, or that $B_2$ is compressed
along with the gas, i.e., $B_2 = \Rtot \, B_0$. 
The actual situation should lie somewhere in between.
We assume the unshocked proton temperature, $\TempPro=10^4 \, \mbox{K} =
\TempElec = \TempHe$, where $\TempElec$ $(\TempHe)$ is the upstream
electron (helium) temperature. Our solutions are insensitive to the
upstream temperature as long as it is below $\sim 10^6$ K.
The shocked electron to proton temperature ratio, $\TempRatio$, is a
free parameter which has not yet been deduced from first 
principles for collisionless shocks.
The electron temperature is an important parameter for thermal \brem\
and X-ray line emission, but is not 
strongly constrained for \GGG\ because no lines are seen. 
In any case, we assume $\TempRatio = 1$ in all of our models but note
that if electron-proton equilibration timescales are long compared to
dynamic timescales, $\TempRatio$ could be considerably less than one,
lowering the thermal \brem\ to \syn\ ratio.\footnote{The
combination of high post-shock density and low post-shock proton
temperature means that electrons should equilibrate on a faster
timescale in shocks undergoing efficient acceleration compared to
those where little acceleration occurs
\citep{DEB00}. 
}
\listromanDE
The injection efficiency, $\etainjP$, is the fraction of thermal
protons that end up as accelerated particles. This parameter
influences the overall acceleration efficiency.
\listromanDE
The electron to proton ratio at relativistic energies, $\epRatio$, is
an arbitrary parameter in our models but is expected to lie
between 0.01 and 0.05 if  $\epRatio$ in Galactic cosmic rays is
typical of that produced by the strong shocks in young SNRs.  
This factor influences the normalization of the photon emission from
electrons and is important for determining the \pion\ contribution to
TeV $\gamma$-rays relative to \IC. 
The maximum energy cosmic rays obtain depends on the
scattering mean free path, $\lambda$, which is assumed to be,
\begin{equation}
\label{eq:mfp}
\lambda = 
\etamfp \, \rgmax \, \left ( \rg / \rgmax\right )^\alpha
\ ,
\end{equation}
where $\etamfp$ is a constant, $\rg=
p/(q B)$ is the gyroradius in SI units, $\rgmax$ is the gyroradius at
the maximum momentum, $\pmax$, and 
$\alpha$ is a constant parameter.  Small values of $\etamfp$ imply
strong scattering and allow higher maximum proton energies in a given
system.  We use the Bohm limit ($\etamfp =1$) in all of our models and
note that while good fits to the observations can be obtained with
larger $\etamfp$, these models will have similar acceleration
efficiencies with {\it lower} proton maximum energies than the models
we show.
\listromanDE
Finally, $\ShellDen$ is the proton number density in the shell
which models the density enhancement presumed to exist in the NW
rim. 
\citet{Slane99} conclude, from the lack of thermal emission, that
the mean density around the remnant is quite low. 
With large uncertainties, they estimate upper limits on the ambient
density of $0.014$--$0.28 \, (\dSNR/6 \, \mbox{\rm kpc})^{-1/2}$
cm$^{-3}$, with a higher upper limit ($< 1 \, (\dSNR/6 \, \mbox{\rm
kpc})^{-1/2}$ cm$^{-3}$) for the northwest rim.

% tttt - 1111
\begin{table}
\begin{center}
\caption{Forward shock parameters for \GGG. All models listed have
$\etamfp = 1$,
$\alpha = 0.5$,
$\TempPro=10^4$ K, and $\dSNR=6$
kpc.  
\vskip12pt
\label{table-1}}
\begin{tabular}{crrrrrrrrrrr}
\tableline
\tableline
Input
&A
&B
&C
&
\\
\tableline
$\EnSN$ [$10^{51}$ erg] 
&2
&3
&4
&
\\
$B_0$ [\muG]
&2.4
&2.4
&16
&
\\
$\etainjP$
&$1\xx{-3}$ 
&$5\xx{-6}$ 
&$1\xx{-3}$ 
&
\\
$\epRatio$
&0.03 
&0.05
&0.04
&
\\
$\ShellDen$ [\pcc]
&0.015
&0.02
&0.008
&
\\
\tableline
Output& & \\
\tableline
$\Vsk$ [\kmps] 
&980
&1100
&2100
&
\\ 
$\tSNR$ [yr]
&9000
&8000
&5000
&
\\ 
$\DelTime$ [yr]
&5000
&4700
&2200
&
\\ 
$\MSZ$ 
&60
&70
&130
&
\\ 
$\MAZ$ 
&27
&35
&6.3
&
\\ 
$\Rtot$ 
&6.1
&4.1
&4.3
&
\\ 
$\Rsub$ 
&3.0
&4.0
&3.4
&
\\
$\gameff$ 
&1.4
&1.65
&1.6
&
\\
$B_2$ [\muG] 
&15
&9.6
&16
&
\\
$k \DStemp$ [keV] 
&$0.33$ 
&$1.2$ 
&$3.4$
&
\\ 
$k \DStp$ [keV]
&$1.0$ 
&$1.2$ 
&$4.5$
&
\\ 
$\EmaxPro$ [TeV]
&14
&16
&70
&
\\ 
$\EmaxElec$ [TeV]
&14
&16
&32
&
\\ 
$\EffRel$ 
&0.55
&0.03
&0.20
&
\\
$\sigma(2\, \mbox{GHz})$
&$-0.45$
&$-0.51$
&$-0.54$
&
\\
\tableline
Flux& & \\
\tableline
$\fvol$
&0.16
&6.8
&0.19
&
\\
\tableline
\end{tabular}
\end{center}
\label{numbers}
\end{table}

The output values are also given in Table~\ref{table-1}.  The shock
speed, $\Vsk$, is the average obtained from the hydro simulation over
the shell thickness between $R=34$ and 40 pc.
The age of \GGG\ is unknown, but \citet{Slane99} have used emission
measures to estimate a range $2000 < \tSNR < 40,000$ yr. The value in
the Table is that necessary to yield an outer shock radius of
$\Rsk \sim 40$ pc consistent with an angular diameter $\sim
45'$ at a distance of 6 kpc.
The time the shock spends between 34 pc and 40 pc is $\DelTime$.
Given $\Vsk$ and the density, temperature, and magnetic field in the shell,
the sonic and \alf\ Mach numbers are determined.
The above parameters allow
%
%\DCE\DCE
%
the nonlinear shock model to determine $\Rtot$, $\Rsub$, and
$\gameff$, as well as the proton, electron, and helium spectra.  Note
that consistent values of $\Rtot$ and $\gameff$ are found by iterating
the hydro simulation with the particle acceleration calculation.
The shocked magnetic field is
$B_2$, and the shocked proton temperature is $\DStemp$. The shocked
proton temperature for the corresponding test-particle shock, $\DStp$,
is also given.

The particle momentum distributions, $f(p)$, are calculated as in
\citet{EBB00} with the turnover at the highest
energies described by:
\begin{equation}
\label{eq:expo}
f(p) \rightarrow
f(p) \,
%\amax \, \left( \frac{p}{\pint} \right)^{-\qmin} \,
\exp{\left [ -\frac{1}{\alpha} 
\left ( \frac{p}{\pmax} \right )^{\alpha}
\right ]}
\ .
\end{equation}
Here, $\alpha$ is the same as in Eqn.~(\ref{eq:mfp})  
and $\pmax$ is determined by setting the
acceleration time equal to $\DelTime$, or by setting the diffusion
length of the highest energy particles equal to 1/10 $\Rsk$,
whichever gives the lowest $\pmax$ \citep[see][]{BaringEtal99}.
It turns out that a relatively slow turnover (\iec\ $\alpha \sim 0.5$)
is required to obtain the proper X-ray slope and to join the radio and
X-ray observations. A similar effect was required to match the
intensity and shape of the X-ray observations in SN1006
\citep[][]{Reynolds98,BKP99,EBB00}, and we
take $\alpha=0.5$ in all of our models.
The electron and proton energies corresponding to $\pmax$ are
$\EmaxElec$ and $\EmaxPro$ respectively and are listed in the Table.
%% As we argued above, nonlinear shocks can be highly efficient and can
%% place a majority of the total energy flux into relativistic and
%% escaping particles (mainly protons).
%
The acceleration efficiency, $\EffRel$, is defined as, 
\begin{equation}
\EffRel =
\frac{\Frel + \Fesc}{(1/2) \rho_0 \Vsk^3}
\ ,
\end{equation}
where $\Frel$ and $\Fesc$ are the fractions of energy flux in
relativistic protons and escaping protons, respectively.

The spectral index at 2 GHz, where $F_\nu \propto \nu^{\sigma}$, is
shown in the next to last row of Table~\ref{table-1}.
Radio data has been collected at 2.4~GHz from \GGG\ but has not yet
been analyzed. When the 2.4~GHz flux is combined with the existing
1.4~GHz flux, the inferred spectral index can be used as a
check of our models and may help constrain parameters.
Finally, the last row in the Table is the fraction of the shell
volume, $\fvol$, producing the emission seen from the NW limb. Based
on equation~(\ref{eq:Emission}), models with $\fvol > 0.25$ are
unacceptable.

\begin{figure}[!hbtp]              % Figure 3
\epsscale{0.6}
\plotone{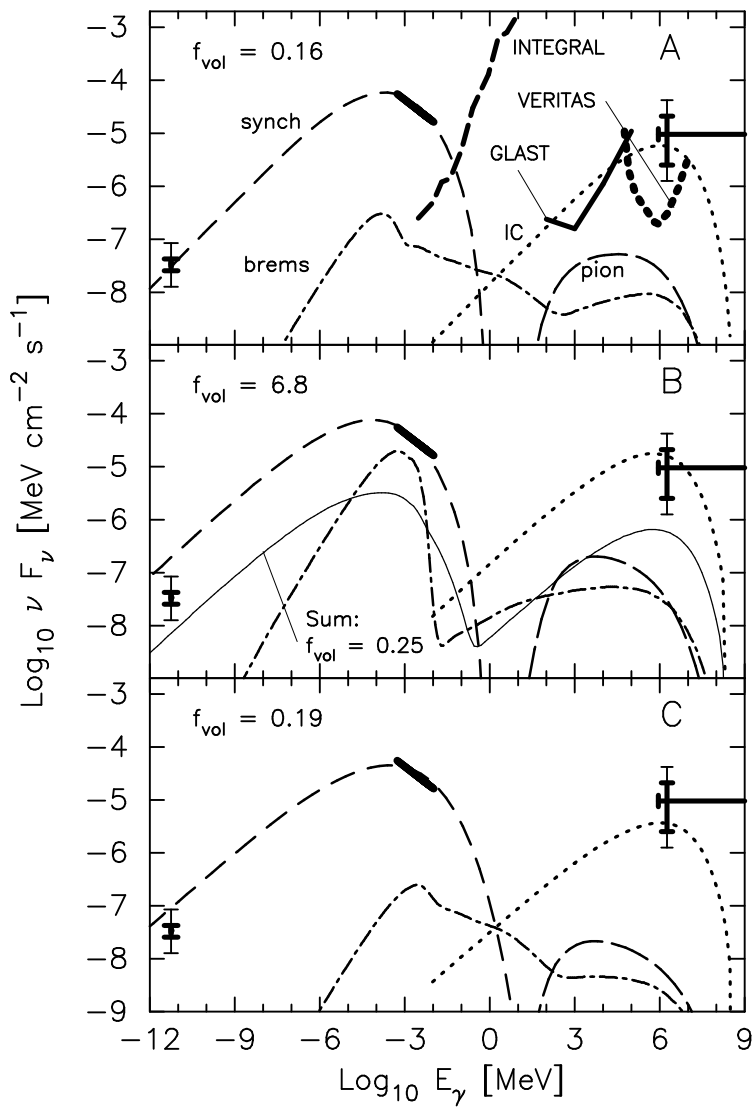}
\figcaption{
Comparison of broad-band emission from G347.3-0.5 with diffusive shock
acceleration models.
Model $A$ uses a likely set of parameters with $B_2= \Rtot \,
B_0$, model $B$ is a fit
obtained with a test-particle shock, and model $C$ was obtained with
parameters chosen to obtain the highest proton energy.
In all panels, the short-dashed curves are \syn, the dot-dashed are
\brem, the dotted are \IC, the long-dashed curves are \pion, and the
normalization is set by matching the \syn\ fluxes at $\sim 5$ keV. The
heavy line at X-ray energies is the fit to the NW rim observations
from \citet{Slane99}.  The gamma-ray point is from
\citet{Muraishi2000}, and the radio data is first published here.  
The light-weight error bars on the radio and TeV points are included
to compensate for the possible difference in relative emission volumes
between the various energy bands.
In
the top panel we show the expected continuum sensitivity limits for
the GLAST spacecraft \citep{Gehrels99}, the INTEGRAL spacecraft
\citep{Winkler96}, and the VERITAS array \citep{Weekes99}.
The solid curve in $B$ is the total emission with $\fvol = 0.25$.
\label{fig:broadband}
}
\end{figure}

\subsection{Fits to \GGG}

The broad-band emission from \GGG\ allows us to constrain the 
parameters of the shock model to fairly narrow ranges.
In the top panel of Figure~\ref{fig:broadband} (model $A$), we show a
satisfactory fit to the broad-band observations
with $B_2= \Rtot \, B_0$.
The parameters for
this model and two others discussed below are listed in
Table~\ref{table-1}.
The four components of the emission are labeled and it's clear that
the \syn\ emission produces the radio and X-ray emission, while the
TeV point is fit with \IC\ photons from electrons scattering off the
2.73~K background radiation.  The value $\fvol = 0.16$ is consistent
with the morphology and $\ShellDen = 0.015$ \pcc\ is consistent with
the upper limit estimate of \citet{Slane99}. We note that using a
lower $\EnSN$ would force a smaller $\ShellDen$ and a larger $\fvol$;
$\EnSN > 1\xx{51}$ erg is clearly favored.

\begin{figure}[!hbtp]              % Figure 4
\epsscale{0.6}
\plotone{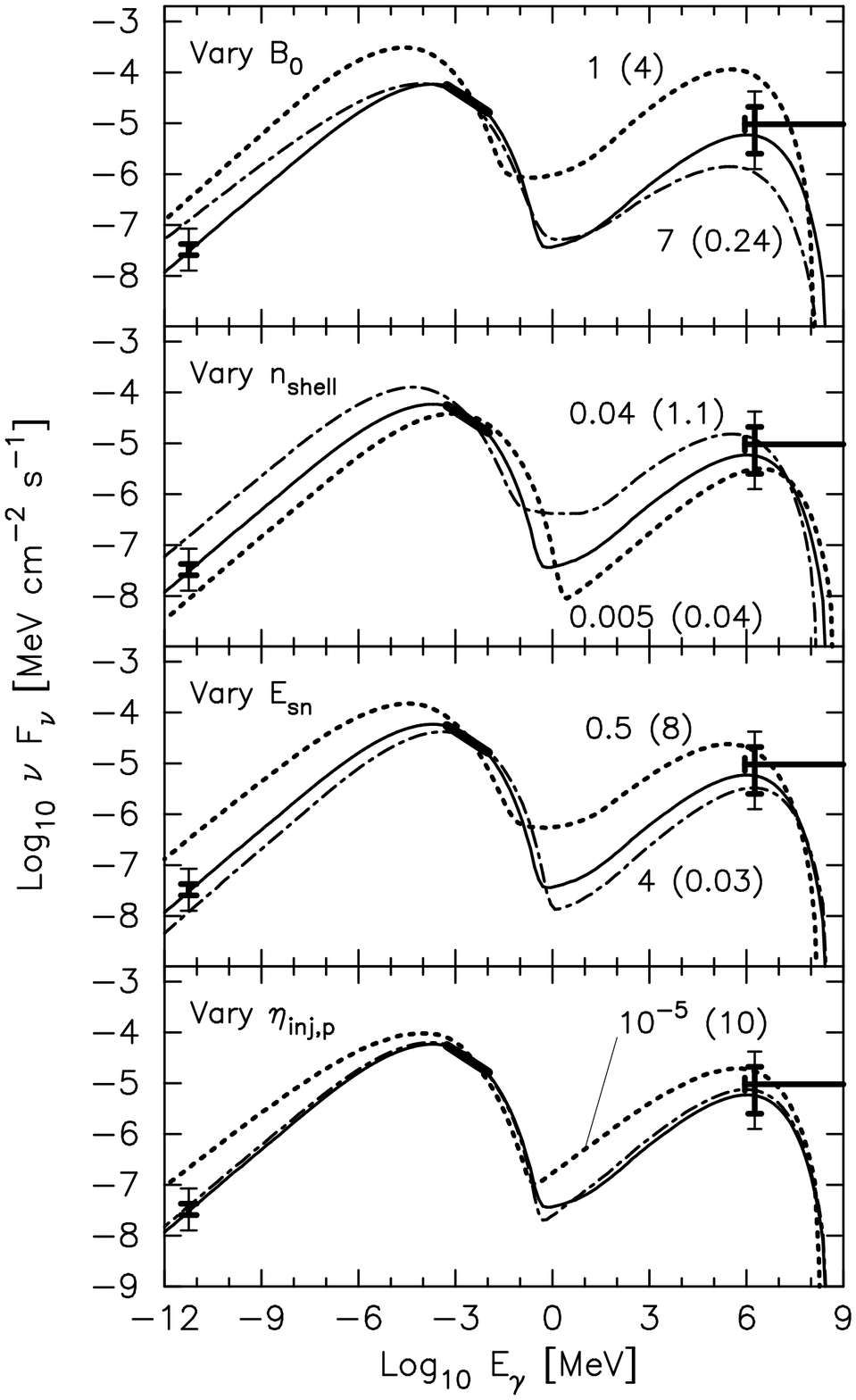}
\figcaption{
These plots and data are as in Figure~\ref{fig:broadband} except here
the curves are the sum of \syn, \brem, \IC, and \pion\ emission.  In
each panel, the labeled parameter is varied as all other parameters
are held fixed with the values shown in Table~\ref{table-1} for model
$A$. The numbers in parentheses are $\fvol$.
These models all have $B_2= \Rtot \, B_0$.  
In the bottom panel, 
$\fvol \simeq 10$ for $\etainjP = 10^{-5}$ (dotted curve),
$\fvol \simeq 0.3$ the $\etainjP = 10^{-4}$ (dot-dashed curve), and
$\fvol \simeq 0.16$ for $\etainjP = 10^{-3}$ (solid curve).
\label{fig:vary}}
\end{figure}

Using model $A$ as a reference, we investigate how sensitive the fit
is to particular parameters. In Figure~\ref{fig:vary} we show
variations in $B_0$, $\ShellDen$, $\EnSN$, and $\etainjP$. In each
panel we keep all parameters except the one labeled fixed to the
values used in model $A$. For all models, we
have normalized the \syn\ emission to the flux in the
X-ray band and plot the sum of the four photon components. The solid
curve in each panel is model $A$.
In the top panel, the dotted curve is for $B_0=1$ \muG, the solid
curve is for $B_0=2.4$ \muG, and the dot-dashed curve is for $B_0=7$
\muG. 
In all cases, the number in parentheses is $\fvol$.
The dependence on $B_0$ is quite strong and the shape of the
broad-band emission clearly sets upper and lower limits on
$B_0$. Furthermore, the $B_0=1$ \muG\ result can be excluded because
$\fvol \simeq 4$; a low $B$ produces too little flux.

The changes produced by varying $\ShellDen$ are shown in the second
panel of Figure~\ref{fig:vary}.  The dotted curve is for $\ShellDen=
0.005$ \pcc, the solid curve is for $\ShellDen= 0.015$ \pcc, and the
dot-dashed curve is for $\ShellDen= 0.04$ \pcc.  Clearly, the
$\ShellDen = 0.005$ \pcc\ result is too flat in the X-ray band, while
the $\ShellDen= 0.04$ \pcc\ result is somewhat too steep. The value
$\fvol\simeq 1.1$ for $\ShellDen= 0.04$ \pcc, shows that this high
density example produces too little flux. This occurs because $\Vsk$
is low in this case (520 \kmps) producing a low $\EmaxElec$ and
forcing a higher normalization to match the high energy X-rays.
%  but could be accounted for by assuming a closer 
% distance and/or a greater supernova explosion energy.

In the third panel, we vary $\EnSN$ from $0.5\xx{51}$ erg (dotted
curve) to $2\xx{51}$ erg (solid curve) to $4\xx{51}$ erg (dot-dashed
curve). The low explosion energy gives an unacceptable fit because of
the shape in the X-ray band and because $\fvol\simeq 8$. However, for
$\EnSN = 4\xx{51}$ erg the shape is still reasonably good considering
the uncertainties in the model and the normalization ($\fvol \simeq
0.03$) while low, cannot be excluded.

In the bottom panel of Figure~\ref{fig:vary} we show results varying
the injection efficiency. The dotted curve is for $\etainjP=10^{-5}$,
the dot-dashed curve is for $\etainjP=10^{-4}$, and the solid curve is
model $A$ with $\etainjP=10^{-3}$. The shape of the sum of the
emission is virtually identical for $\etainjP=10^{-4}$ and
$10^{-3}$. However, the intensity of the emission varies with
$\etainjP$ and for $\etainjP=10^{-4}$, $\fvol \simeq 0.3$, while for
$\etainjP=10^{-3}$, $\fvol \simeq 0.16$. The constraint that the
emission is only observed from a small fraction of the shell implies
that $\etainjP > 10^{-4}$. This will be true unless $\EnSN$ is
considerably greater than $2\xx{51}$ erg.
The $\etainjP=10^{-5}$ result, which is what would be
expected if the shock {\it did not} accelerate particles efficiently,
is unacceptable because of the overall shape and low emissivity (i.e.,
$\fvol \simeq 10$). 

To further investigate the limits on the acceleration efficiency, we
choose $\etainjP$ low enough so only a few percent of the total energy
flux ends up in relativistic particles and freely adjust other
parameters to obtain a fit.
With $\etainjP=5\xx{-6}$ (yielding $\EffRel \sim 3\%$) we are able
to obtain an approximate match to the shape of the observed emission
with Model $B$ in Figure~\ref{fig:broadband}
and Table~\ref{table-1}.
Note that this model requires $\EnSN = 3\xx{51}$ erg; lower energies
give poorer fits.
% by increasing
% $B_0$ from 1.8 to 3 \muG, and increasing $\epRatio$ from 0.01 to 0.05
% compared to model $A$.  These results are shown as model $C$ in
% Table~\ref{table-1} and Figure~\ref{fig:broadband}.
%
While the shape of the continuum emission is a reasonable match to the
observations, the normalization requires $\fvol \sim 6.8$, well
above an acceptable value.
Furthermore, 
the \syn\ emission from the test-particle shock is
somewhat too steep to simultaneously match the radio and X-ray fluxes,
and
%
% and no combination of parameters will allow a better match as long as
% the acceleration efficiency is low.  Furthermore, since $\fvol=1.7$
% when we assume $\dSNR=6$ kpc, a closer distance and/or
% larger explosion energy is required to match the overall flux level.
% Perhaps a more important difference between models $A$ and $C$ is 
% 
the ratio of \syn\ to \brem\ continuum emission in the X-ray range is
less than a factor of 3.
% In
% $C$, the \brem\ to \syn\ ratio is much greater and a lower value of
% $\etainjP$ would increase this ratio.
%
If we further decrease $\etainjP$, the \syn\ to \brem\ ratio will drop
and at some point
%
% If we had included 
X-ray emission lines would stand
above the continuum flux in conflict with
the observations. 
This conclusion depends,
of course, on $\TempRatio$ and could be weakened by assuming
$\TempRatio \ll 1$. However, the relative acceleration efficiency of
electrons and protons is also likely to scale as $\TempRatio$
\citep[][]{EBB00} making it difficult to produce the observed radio
and TeV flux if $\TempRatio$ is too small.
Furthermore, if a flatter \syn\ slope is required to give a better
match between the radio and X-rays (as in model $A$), this forces the
shock to be more efficient (to produce a larger $\Rtot$) and implies a
larger $\etainjP$.  The overall flux, the lack of X-ray emission
lines, and the radio to X-ray flux ratio all point toward efficient
shock acceleration.
Using this model we can get a firm lower
limit on the acceleration efficiency by increasing $\etainjP$ until
$\fvol \sim 0.25$.  We find $\fvol \lesssim 0.25$ for $\etainjP >
3\xx{-5}$, yielding a lower limit $\EffRel > 0.25$, i.e., even with
$\EnSN = 3\xx{51}$ erg, the shock must place at least 25\% of its
kinetic energy flux in relativistic particles.
Note that we took $\epRatio = 0.05$ for this example. A lower
$\epRatio$ would imply a higher acceleration efficiency.

%7 Another difference between models $A$ and $C$ lies in the importance
%7 of the \pion\ emission relative to \IC. While \IC\ dominates in either
%7 case, the \pion\ contribution is considerably greater in model $A$
%7 where $\etainjP$ is large and $\epRatio$ is small.
%7 
%7 We have investigated the effects produced by a presupernova wind
%7 ($s=2$) where $A_0$ replaces $\ProDenUpS$ as an important parameter
%7 and find they are small. The inferred values of $\Vsk$, $\Rsk$,
%7 $B_0$, etc., are given in Table~\ref{table-1} for model $E$ assuming
%7 $B_2= \Rtot \, B_0$. All values are similar to model $A$ as is the fit
%7 which is not shown.

%
In order to investigate the maximum proton energies that \GGG\ can
generate, we have produced Model $C$ (bottom panel of
Figure~\ref{fig:broadband}) where we have chosen parameters (including
taking $B_2 = B_0$) with the intention of maximizing $\EmaxPro$.  
Very generally, diffusive shock acceleration yields the same maximum
energy per {\it charge} for all particles
\citep[\egc][]{Drury83,BaringEtal99}, so the only way the proton and
electron spectra can have different maximum energies is for losses
(\syn\ and/or \IC) to be important for the electrons.
Unless we have neglected an abundant supply of photons in addition to
the 2.73~K background radiation, inverse-Compton losses are not
important and only when the magnetic field is large can $\EmaxPro$ be
much greater than $\EmaxElec$.
However, a large field produces a lower \alf\ Mach number which
weakens the acceleration and, in order to fit the broad-band spectrum,
a higher shock speed is required.  This can be achieved with a larger
$\EnSN$ and/or a lower $\ShellDen$, and we have taken $\EnSN =
4\xx{51}$ erg and $\ShellDen = 0.008$ \pcc\ for Model $C$.
The observed
radio flux sets the strength of the magnetic field in the emission
region combined with the relativistic electron density, i.e., the flux
is loosely proportional\footnote{This is not a strict proportionality
because changing either $B$ or the density produces changes in the
acceleration model and/or the SNR evolution.}
to $B_2^{3/2} \, n_e$.
Furthermore, the TeV \IC\ emission comes from
the same electrons that produce the X-rays. Since the \IC\ emission is
independent of $B$, it constrains the number of relativistic electrons.
The combination of these constraints allows us to limit 
$\EmaxPro$ and we find in our most optimistic case that $\EmaxPro
\lesssim 70$ TeV.

\begin{figure}[!hbtp]             % Figure 5
\plotone{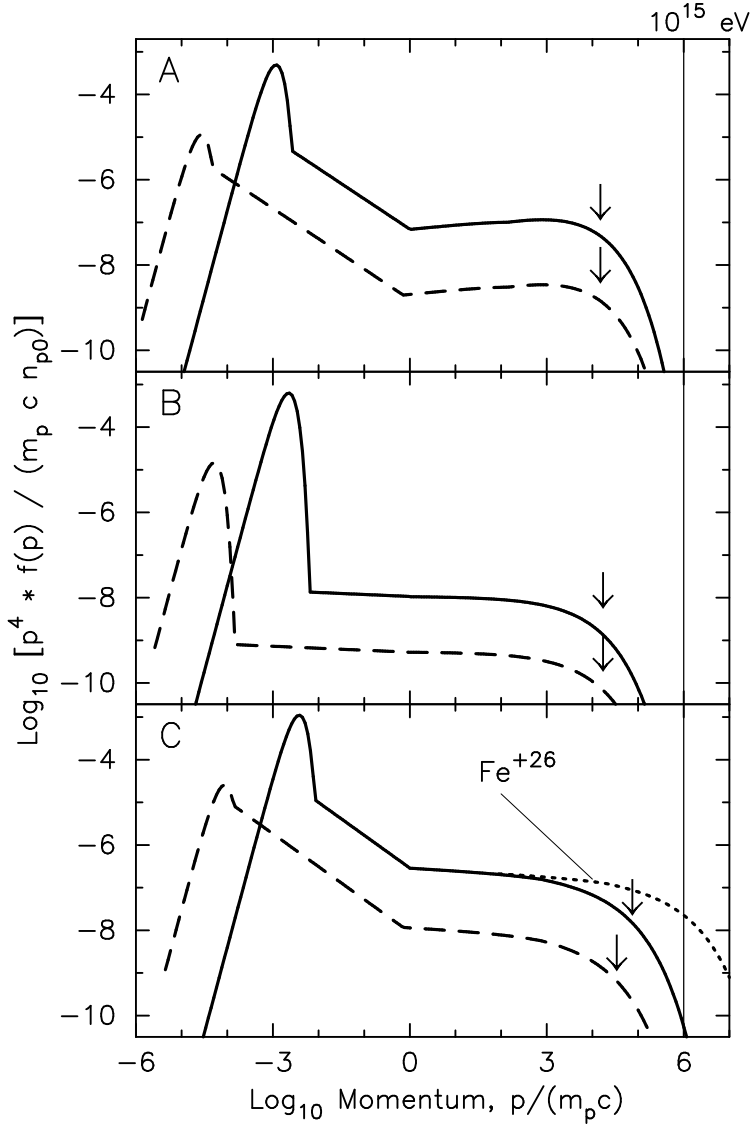} \figcaption{ Proton (solid curves) and electron
spectra (dashed curves) for models $A$, $B$, and $C$ used to generate
the photon spectra shown in Figure~\ref{fig:broadband}. The arrows
indicate $\EmaxPro$ and $\EmaxElec$. The dotted curve in the bottom
panel shows the high momentum portion of a possible Fe$^{+26}$
spectrum from Model $C$. The normalization relative to protons is that
shown in Figure~8 of \citet{EDM97}.
\label{fig:spectra}}
\end{figure}

In Figure~\ref{fig:spectra} we show the momentum distribution
functions, $f(p)$, for models $A$, $B$, and $C$, where we have plotted
the dimensionless quantity, $p^4 \, f(p) /(m_p \, c \, \ProDenUpS)$,
to emphasize the spectral curvature.
In this representation, the test-particle result of $f(p) \propto
p^{-4}$ is a horizontal line (model $B$).  The concave shape of $f(p)$
(at superthermal energies and below the cutoff) is evident for models
$A$ and $C$, as is the large difference between the relativistic
particle populations in these shocks with efficient acceleration
compared to the inefficient shock $B$.  As noted above, a flat
electron spectrum (at energies greater than $m_p \, c^2$) produced by
a large $\Rtot$ ($\Rtot \simeq 6$ in model $A$) allows a good match
between the radio and X-ray observations.

The arrows in Figure~\ref{fig:spectra} indicate the positions of
$\EmaxPro$ and $\EmaxElec$ as determined from equation~(\ref{eq:expo})
with $\alpha = 0.5$. In all cases, $f(p)$ starts to fall off
exponentially at momenta less than $\pmax$ and much below $p=10^6 \,
m_p\, c$.
Furthermore, if G347.3-0.5 continues to evolve uniformly with no
magnetic field amplification, it will not
produce higher energy particles as it ages \citep[for a more complete
discussion of the time particles take to obtain their maximum energies,
see][]{BerezV97,BerezV00}.
Since the causes of magnetic field amplification are still uncertain,
it remains possible that higher fields and therefore higher energy
particles could be produced in the future.
Magnetic field amplification apparently occurs in Cas A
\citep[\egc][]{Gull75,Keohane98} but if it is due to Rayleigh-Taylor
instabilities in this young remnant as commonly assumed, it is
unlikely that this mechanism would occur in \GGG, which is much older.

\section{Conclusions}

G347.3-0.5 is one of a growing number of shell-type SNRs showing
nonthermal X-rays believed produced by \syn\ emission from TeV
electrons accelerated by the remnant shocks.  In an attempt to have a
consistent, broad-band spectrum, we have combined recent X-ray and TeV
$\gamma$-ray observations of the remnant with new radio observations
and have restricted all of the observations to the bright northwest
rim of the shell.
Comparing a spherically symmetric, wind-shell model of SNR
evolution and efficient diffusive shock acceleration against the
observations shows that the observed emission can be well fit across
$\sim 18$ decades in energy with a single set of parameters all close
to expected supernova and ISM values.
The densities we find for the emission region ($\sim 0.01$ protons
cm$^{-3}$) are consistent with the upper limits found earlier by
\citet{Slane99}, and are below those expected for either a molecular
cloud or a swept-up shell produced as the pre-SN wind interacts with
the interstellar medium (or with a slower red supergiant wind).  This
suggests the shock is still within the low density bubble but has
encountered a density enhancement perhaps caused by perturbations in
the pre-SN wind.

We put limits on the acceleration efficiency and find that \GGG\ is
currently putting at least $25$\% (and more likely $\sim 50$\%) of the
forward shock ram energy flux into relativistic particles (\iec\
cosmic rays).
Recently, \citet{Hughes00b} have inferred efficient shock acceleration
in SNR E0102.2-7219 in the Small Magellanic Cloud from the shock
speed, determined from expansion measurements, and the postshock
electron temperature, determined from X-ray line models (unlike \GGG,
lines are visible above the \syn\ continuum in the post-forward shock
gas).
They concluded that the extremely low postshock
temperature can only be reconciled with the high shock speed if a
sizable fraction ($\sim 50$\%) of the shock energy goes into cosmic
rays. Our results are fully consistent with this and with the
important point that efficient particle acceleration causes a strong
coupling between the broad-band emission from relativistic particles
and the properties of the shocked heated X-ray emitting gas
\cite[see][]{DEB00}.
We have verified that Sedov models (not shown) with no density
enhancement produce satisfactory nonlinear fits to the observed
spectral shape with input parameters similar to those given here, but
are inconsistent with the observed remnant diameter and
distance.\footnote{In standard Sedov models, the relatively high
shock speeds and ambient densities which produce acceptable fits to
the observed shape occur at shock radii which are too small to be
consistent with a remnant diameter of $45'$ (or even $30'$) at $\dSNR
\sim 6$ kpc.  Only if $\dSNR < 3$ kpc could standard models produce
acceptable fits. The wind-shell model allows the shock to propagate
for large distances in the low density wind before encountering the
higher density emission region.}
We conclude that the nonlinear aspects we infer, in particular the
required high acceleration efficiency, do not depend significantly on
the geometric details of the wind-shell model while the overall
normalization does.
%
%% On the other hand, if
%% a sizable fraction of the flux we see comes from regions, such as
%% dense knots, which are not acceleration sites, our results could
%% change substantially.

Furthermore, we find that the maximum energies of the cosmic-ray
electrons and protons are well below $10^{15}$ eV.  While it has been
known for some time that SNRs emitting nonthermal X-rays do not
produce {\it electrons} to energies near $10^{15}$ eV
\citep[\egc][]{RK99}, the nonlinear effects we model make it possible
%
% our model is able 
to infer the proton distribution from photons emitted by
electrons.
For no magnetic field compression in the shock and other parameters
adjusted to produce the highest possible $\EmaxPro$ and still give a
reasonable fit to the observations, we find
%% Our most conservative estimate, which assumes no magnetic field
%% compression in the shock, gives 
%
a proton maximum energy $\lesssim 70$ TeV.
More realistic parameters with magnetic field compression
yield $\EmaxPro \sim 20$ TeV.
However, SNR shocks will also accelerate heavier ion species such as
He, C, O, and Fe, and the maximum energy in diffusive shock
acceleration scales as charge. Therefore, this SNR could produce a
power-law Fe$^{+26}$ spectrum to $\sim 10^{15}$ eV. To illustrate, we
have estimated the Fe$^{+26}$ spectrum produced by our model $C$ and
placed it
%
%% , with {\it arbitrary normalization}, 
in the bottom panel of Figure~\ref{fig:spectra}.
The normalization is determined by the cosmic abundance of iron ($\sim
3.1\xx{-5}$ iron nuclei for every proton) plus the enhancement
expected because iron is injected and initially accelerated when it is
trapped in dust grains. In fact, the normalization of iron relative to
hydrogen in Figure~\ref{fig:spectra} is just that shown in Figure~8 of
\citet{EDM97} at $10^8$ keV/nucleon.
Cosmic-ray observations in the
$10^{14}-10^{16}$ eV range typically measure the total energy of the
particles and have large uncertainties in differentiating individual
ion species
\citep[\egc][]{Bern98,Glasmacher99,Swordy00}. Nevertheless, a number
of authors have produced models of this transition energy range adding
together the fluxes from different ion species with cutoffs that scale
as charge \citep[\egc][]{Stanev93,EW00}.  Our high $B_0$ results are
in general agreement with this work.

\acknowledgments The authors thank J. Blondin and the referee
A. Decourchelle for helpful comments.  This work was supported in part
by the National Aeronautics and Space Administration through contract
NAS8-39073 and grant NAG5-4803.
The Australia Telescope is funded by the Commonwealth of Australia for
 operation as a National Facility managed by CSIRO.
 B.M.G. acknowledges the support of NASA through Hubble Fellowship
 grant HF-01107.01-98A awarded by the Space Telescope Science
 Institute, which is operated by the Association of Universities for
 Research in Astronomy, Inc., for NASA under contract NAS 5--26555.

\newcommand{\aaDE}[3]{ 19#1, A\&A, #2, #3}
\newcommand{\aatwoDE}[3]{ 20#1, A\&A, #2, #3}
\newcommand{\aasupDE}[3]{ 19#1, {\itt A\&AS,} {\bff #2}, #3}
\newcommand{\ajDE}[3]{ 19#1, {\itt AJ,} {\bff #2}, #3}
\newcommand{\anngeophysDE}[3]{ 19#1, {\itt Ann. Geophys.,} {\bff #2}, #3}
\newcommand{\anngeophysicDE}[3]{ 19#1, {\itt Ann. Geophysicae,} {\bff #2}, #3}
\newcommand{\annrevDE}[3]{ 19#1, {\itt Ann. Rev. Astr. Ap.,} {\bff #2}, #3}
\newcommand{\apjDE}[3]{ 19#1, {\itt ApJ,} {\bff #2}, #3}
\newcommand{\apjletDE}[3]{ 19#1, {\itt ApJ,} {\bff  #2}, #3}
\newcommand{\apjlettwoDE}[3]{ 20#1, {\itt ApJ,} {\bff  #2}, #3}
\newcommand{\apjpress}{{\itt ApJ,} in press}
\newcommand{\apjletpress}{{\itt ApJ(Letts),} in press}
\newcommand{\apjsDE}[3]{ 19#1, {\itt ApJS,} {\bff #2}, #3}
\newcommand{\apjsubDE}[1]{ 19#1, {\itt ApJ}, submitted.}
\newcommand{\apjsubtwoDE}[1]{ 20#1, {\itt ApJ}, submitted.}
\newcommand{\appDE}[3]{ 19#1, {\itt Astroparticle Phys.,} {\bff #2}, #3}
\newcommand{\apptwoDE}[3]{ 20#1, {\itt Astroparticle Phys.,} {\bff #2}, #3}
\newcommand{\araaDE}[3]{ 19#1, {\itt ARA\&A,} {\bff #2},
   #3}
\newcommand{\assDE}[3]{ 19#1, {\itt Astr. Sp. Sci.,} {\bff #2}, #3}
\newcommand{\icrcsaltlake}[2]{ 1999, {\itt Proc. 26th Int. Cosmic Ray Conf.
    (Salt Lake City),} {\bff #1}, #2}
\newcommand{\icrcsaltlakepress}[2]{ 19#1, {\itt Proc. 26th Int. Cosmic Ray Conf.
    (Salt Lake City),} paper #2}
\newcommand{\jgrDE}[3]{ 19#1, {\itt J.G.R., } {\bff #2}, #3}
\newcommand{\mnrasDE}[3]{ 19#1, {\itt M.N.R.A.S.,} {\bff #2}, #3}
\newcommand{\mnrastwoDE}[3]{ 20#1, {\itt M.N.R.A.S.,} {\bff #2}, #3}
\newcommand{\natureDE}[3]{ 19#1, {\itt Nature,} {\bff #2}, #3}
\newcommand{\phyreptsDE}[3]{ 19#1, {\itt Phys. Repts.,} {\bff #2}, #3}
\newcommand{\revgeospphyDE}[3]{ 19#1, {\itt Rev. Geophys and Sp. Phys.,}
   {\bff #2}, #3}
\newcommand{\rppDE}[3]{ 19#1, {\itt Rep. Prog. Phys.,} {\bff #2}, #3}
\newcommand{\ssrDE}[3]{ 19#1, {\itt Space Sci. Rev.,} {\bff #2}, #3}

\end{document}